\documentclass[unknownkeysallowed, letterpaper]{article}
\usepackage[utf8x]{inputenc}
\usepackage[english]{babel}
\usepackage{amsmath, amssymb, amsthm, amsfonts, latexsym}
\usepackage[affil-it]{authblk}
\usepackage[top=0.5cm,bottom=1.5cm,left=2cm,right=2cm ]{geometry}
\usepackage{subcaption}
\usepackage{graphicx}
\usepackage{float}
\usepackage{appendix}
\usepackage{hyperref}

\title{Team assembly mechanisms and the knowledge produced in the Mexico's National Institute of Geriatrics: a network analysis and agent-based modelling approach}
\author{ Carmen Garc\'ia-Pe\~na $^2$, Luis Miguel Guti\'errez-Robledo $^2$, Augusto Cabrera-Becerril $^3$ and David Fajardo-Ortiz $^1$\footnote{David Fajardo-Ortiz (Corresponding author)\\
Centro de Investigaci\'on en Pol\'iticas, Poblaci\'on y Salud, Universidad Nacional Aut\'onoma de M\'exico, Mexico City, M\'exico. \\ davguifaj@gmail.com}}
\affil{1. Centro de Investigaci\'on en Pol\'iticas, Poblaci\'on y Salud, Universidad Nacional Aut\'onoma de M\'exico, M\'exico City, M\'exico \\2. National Institute of Geriatrics, Mexico City, M\'exico, \\3. Facultad de Ciencias, Universidad Nacional Aut\'onoma de M\'exico, Mexico City, Mexico}
\date{\today}

\begin{document}
\maketitle

\begin{abstract}
Mexico's National Institute of Geriatrics (INGER) is the national research center of reference for matters related to human aging. INGER scientists perform basic, clinical and demographic research which may imply different scientific cultures working together in the same specialized institution. In this paper, by a combination of text mining, co-authorship network analysis and agent-based modeling we analyzed and modeled the team assembly practices and the structure of the knowledge produced by scientists from INGER. Our results showed a weak connection between basic and clinical research, and the emergence of a highly connected academic leadership. Importantly, basic and clinical-demographic researchers exhibited different team assembly strategies: Basic researchers tended to form larger teams mainly with external collaborators while clinical and demographic researchers formed smaller teams that very often incorporated internal (INGER) collaborators. We showed how these two different ways to form research teams impacted the organization of knowledge produced at INGER. Following these observations, we modeled, via agent-based modeling, the coexistence of different scientific cultures (basic and clinical research) exhibiting different team assembly strategies in the same institution. Our agent model successfully reproduced the current situation of INGER. Moreover, by modifying the values of homophily we obtain alternative scenarios in which multidisciplinary and interdisciplinary research could be done.
\end{abstract}
\newpage
\section{Introduction}

The National Institute of Geriatrics (INGER), founded in 2009 in Mexico, became one of the National Institutes of Health of the Ministry of Health in 2012, as an answer to the challenge posed by the ageing of the Mexican population.\cite{1} INGER is the national research center of reference for matters related to  human aging.\cite{1}  Aeging is a complex multidimensional problem that would require the development of interdisciplinary research in order to satisfy the demand of knowledge that provide solutions to the several health problems related to aeging.\cite{2}-\cite{4}.  However, according to its official website \cite{5} and its specific organization manual,\cite{6} INGER exhibits a traditional departmentalized organization in which  research is mainly performed by three departments: basic research, clinical epidemiology and demographic epidemiology. It is important to point out that this is the common model for Mexican institutions in which medical research is performed. \cite{7,8}

Currently, one of the biggest challenge in medical research is the translation of basic research discoveries into clinical practice and societal outcomes. However, this is not an easy task as clinical and basic are quite different research cultures with different jargon research instruments and standards, of scientific quality. \cite{9} One of the key elements of the knowledge translation challenge is the assembly of interdisciplinary (translational) research teams. \cite{10} Importantly, team assembly mechanisms have been described as fundamental elements that determinate the structure of the collaboration networks and the performance of the teams. \cite{11} This model considers three parameters: “team size, the fraction of newcomers in new productions, and the tendency of incumbents to repeat previous collaborations.” \cite{11} More recently, Bakshy, E. and Wilensky developed an agent-based model of team assembly dynamics \cite{12} which is based on the previous work of Guimera et al. \cite{11} However, this model does not consider the coexistence of different research cultures, like clinical and basic research, which could have different team assembly practices and, more importantly, their members could be reluctant to collaborate with scientist outside their disciplinary field. 

In this investigation, we set out to analyze and model the team assembly dynamics of INGER researchers and their impact in the knowledge produced by the institution.

\section{Methodology}

The present study used as a source of information the scientific production (A total of 178 papers published between 2012 and 2017) of 21 INGER scientists (3 officeholders and 5 junior and 13 senior scientists) and 8 former INGER researchers. By a combination of text mining and co-authorship network analysis we analyzed the relation between team assembly practices and knowledge produced at INGER. Also, we developed an agent-based model that allowed us to recreate the dynamics of team assembly at INGER and visualize alternative scenarios. We followed the next steps:

\begin{enumerate}
\item KH Coder, a software for quantitative content analysis (text mining), \cite{13} was used to perform bimodal (authors and terms) correspondence analysis of terms contained in the title and abstracts of the papers. Correspondence analysis is an explorative multivariate technique that provides information on the data structure by summarizing a set of data in two-dimensional graphical form. In the plot, the closer are the variables (authors and terms) the more similar they are.
\item A biomodal network model of papers and authors (INGER researchers and their collaborators) was built and visualized using Cytoscape, an open access software for visualization and analysis of networks.\cite{14} Papers in the bimodal network model were classified as basic research if their content were about phenomena at cellular or biomolecular level whereas papers classified as clinical-demographical research are studies at individual or population level. Researchers were classified according to the department they belong as it is noted on the INGER official website except if they are former INGER researchers or officeholder.
\item The bimodal network model was transformed in a collaboration network of authors which was then divided in years so the research teams of each year could be visualized. The average of team size and the proportion of internal and external collaborators was obtained for basic and clinical research teams.
\item An agent-based model of team assembly\cite{12} was previously build in NetLogo\cite{15} by Bakshy and Wilensky which in turn was based on the work of Guimera et al about the mechanism of team assembly that could determine the collaboration network structure and performance.\cite{11} This model does not consider the coexistence of communities (scientific cultures) that exhibit different team assembly strategies like biomedical and clinical researchers at INGER. Therefore, we modified the original model in order to have two coexisting breeds of scientists (basic and clinical researchers) whose team assembly strategies (The probabilities of choosing incumbents and previous collaborators) can be modulated independently. Moreover, a new system variable, homophily, was added. In our model, homophily is the tendency of scientist to collaborate with members of the same scientific culture (basic and clinical research). By modulating homophily and team assembly strategies, difference scenarios can be simulated. It is important to mention that in order to provide robustness to out model we added randomness to the variables homophily and team size.  The model, which can be run in the open source software (GNU General Public License) NetLogo, the code and a more detailed description can be downloaded here:\url{http://modelingcommons.org/browse/one\_model/5676}
\item Three virtual experiments were run 100 times. Each experiment corresponds to different percentages of homophily (14\%, 46\% and 79\%) in order to generate multidisciplinary and interdisciplinary scenarios. The three experiments reproduce the INGER scenario in terms of team size and probabilities of choosing incumbent (Internal) and new collaborators (external) collaborators. That is, the experiments respond to the hypothetical question of what would happen if INGER researchers would have different levels of homophily. Each experiment last 1500 steps (ticks) and the evolution of the team size and composition of team were plotted. 
\end{enumerate}

\section{Results}

\subsection{Content of INGER papers and the structure of the collaboration network}

The correspondence analysis performed on the title and abstract of the 21 current INGER researchers generated two components (dimensions 1 and 2) that, together, explain just 33.5 per cent of the relations among variables (terms in title and abstract). However, the analysis provides useful information on the organization of the research activities at INGER. The correspondence analysis ({\bf Figure 2}) displayed clinical (from individual to population level) researchers forming a cluster near the origin while basic (sub-individual level) researchers are scattered in the plot. Moreover, the plot shows that INGER papers have been by far mostly clinical-epidemiological research over the years. 2014 is the year, according to the correspondence analysis, with more proportion of basic research papers.  On the other hand, the co-occurrence network of terms in the title and abstract of INGER papers showed a primacy of demographic-epidemiological terms ({\bf Figure 1}). 

On the other hand, the bimodal network of authors and publications of INGER confirmed the results of the correspondence analysis: clinical and demographic researchers are strongly connected each other at INGER while basic researchers are scattered ({\bf Figure 2}). As a matter of fact, biomedical researchers at INGER are more strongly connected to external authors than INGER authors ({\bf Figure 2}). Another important observation is that there is a well-defined leadership among clinical researchers. That is, the two most productive authors at INGER are also the most collaborative and they are located at the core of the network model.

\subsection{Clinical and basic research team assembly profiles}

 The annual slices of the collaboration network (2012-2017), showed that clinical-demographic researchers and basic researchers had different team assembly practices ({\bf Figure 3}).  Clinical-demographic teams tended to be smaller than basic teams (The average team size was 4.78 and 7.48, respectively) whereas basic teams showed a smaller proportion of internal collaborators than clinical researchers (The average proportion of internal collaborators per team was 0.22 and 0.45, respectively) As a matter of fact, just 4\% of basic teams included at least two INGER researchers whereas 43 \% of clinical-demographic teams where formed by at least two INGER researchers.

\subsection{Agent-based modeling of INGER team assembly dynamics and alternative scenarios of multidisciplinary and interdisciplinary research. }

We successfully developed an agent-based model of assembly of interdisciplinary teams which can be run in the open source software (GNU General Public License) NetLogo, the code and a more detailed description can be downloaded here:  \url{http://modelingcommons.org/browse/one\_model/5676} 

The virtual experiments that corresponded to different percentages of homophily successfully reproduced the INGER current stage (14\% of homophily; {\bf Figure 4}) and generated multidisciplinary (Homophily 46\%; {\bf Figure 4}) and interdisciplinary (homophily 79\%; {\bf Figure 4}) scenarios. The three scenarios exhibited stability over the time (steps of ticks) in terms of number of teams with size over the selected team size and the compositions of the teams (basic and clinical members; {\bf Figure 5}). In the case of the simulated INGER it can be observed a giant component of clinical researchers and dispersed teams of basic researchers ({\bf Figure 4 and 5}). With a homophily of 46\%, emerged a giant component that were formed by two clusters of clinical and basic researchers that were united by a small region ({\bf Figure 4 and 5}). This is quite similar to a multidisciplinary research scenario. Finally, with a homophily of 79\% clinical and basic researchers were blended in a giant component ({\bf Figure 4 and 5}). 

\section{Discussion}

As far as we know, this is the first time that a combination of text mining, network analysis and agent-modeling is used to provide a systemic view of the current situation of a research institution (INGER) and to generated alternative scenarios. This combination of analysis tools and modeling can be a fundamental element for the knowledge management of research institutions. By modulating one of the system variables, homophily, while keeping fixed the others the different behaviors could be obtained which fit well with the definitions and empirical observations of multidisciplinary) and interdisciplinary (translational) research. See \cite{ 16, 17} However, our agent-based model of  multi-interdisciplinary research can be adapted to a plurality of medical research institutions like the rest of the National Health Institutes of Mexico, which share a similar institutional design,\cite{7, 8} by modulating the full set of  variables. Moreover, this set of methodologies can be applied to the analysis and management of intramural research performed in the United States National Health Institutes.
As a possible extension of our model would be the inclusion of a third class of researchers, tanslational researchers that could establish the connection between basic and clinical research. Currently, we are working in the building of a model that consider the coexistence of several research communities.
Finally, it is important to mention that our results showed how team assembly strategies could impact the knowledge produced by a research institution (See figures 1 and 2). This could be explained by the phenomenon of homophily whose “pervasive fact of homophily means that cultural, behavioral, genetic, or material information that flows through networks will tend to be localized. Homophily implies that distance in terms of node information translates into network distance.”\cite{18} Therefore, a good management of team assembly practices could lead a better quality of the produced knowledge.

\section{Conclusions}

The combination of data science and agent-based modeling can be a fundamental tool for understanding of the current situation of research institutions and to project possible scenarios. Therefore, this set of methodologies are key source of information for strategic decision making.

\begin{figure}[H]
  \centering
\includegraphics[scale=0.75
]{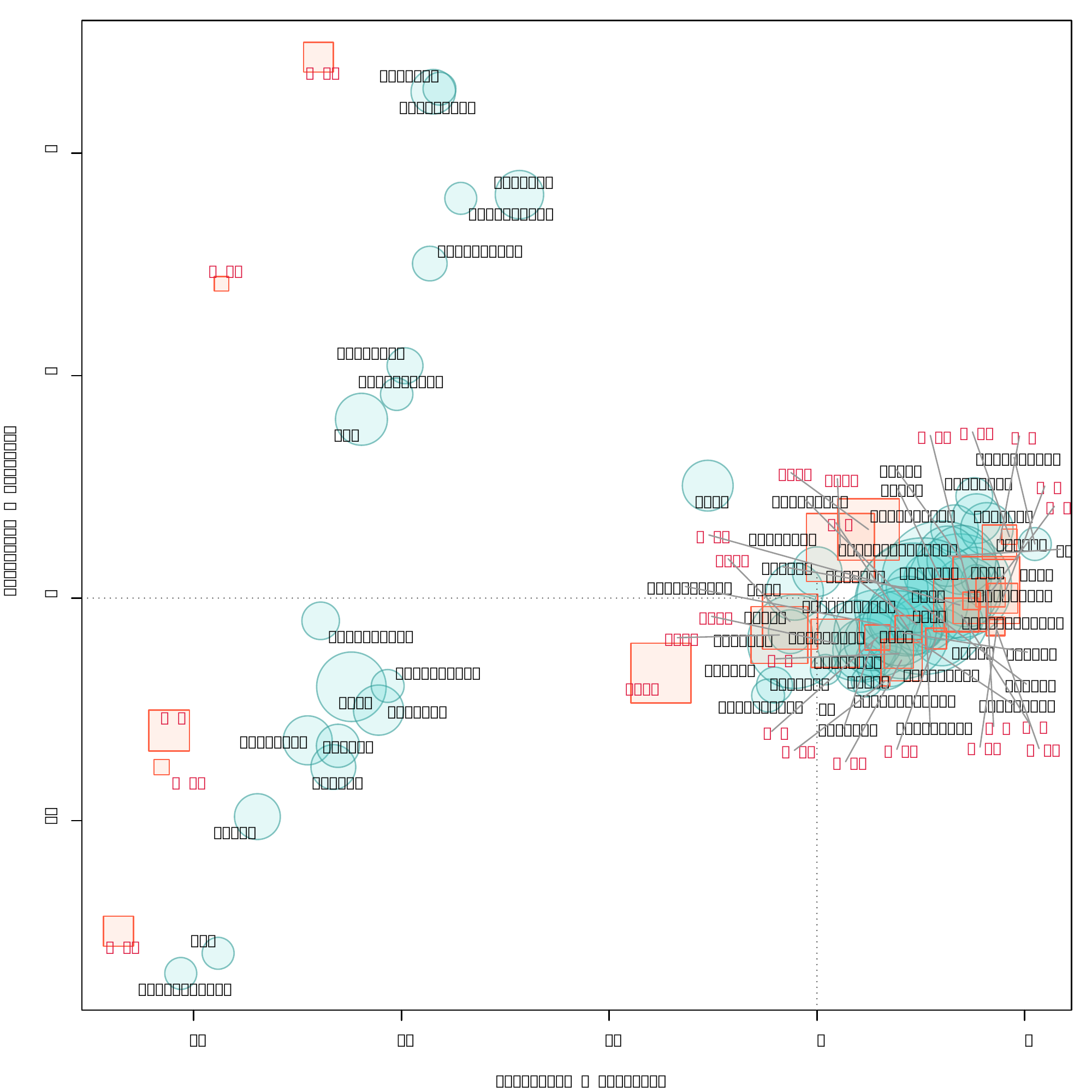}
  \caption{Correspondence analysis of terms in title and abstract of INGER papers. The correspondence plot is intended to graphically show the similarity relation among the variables (Terms, authors and years).}
\end{figure}

\begin{figure}[H]
  \centering
\includegraphics[width=\textwidth]{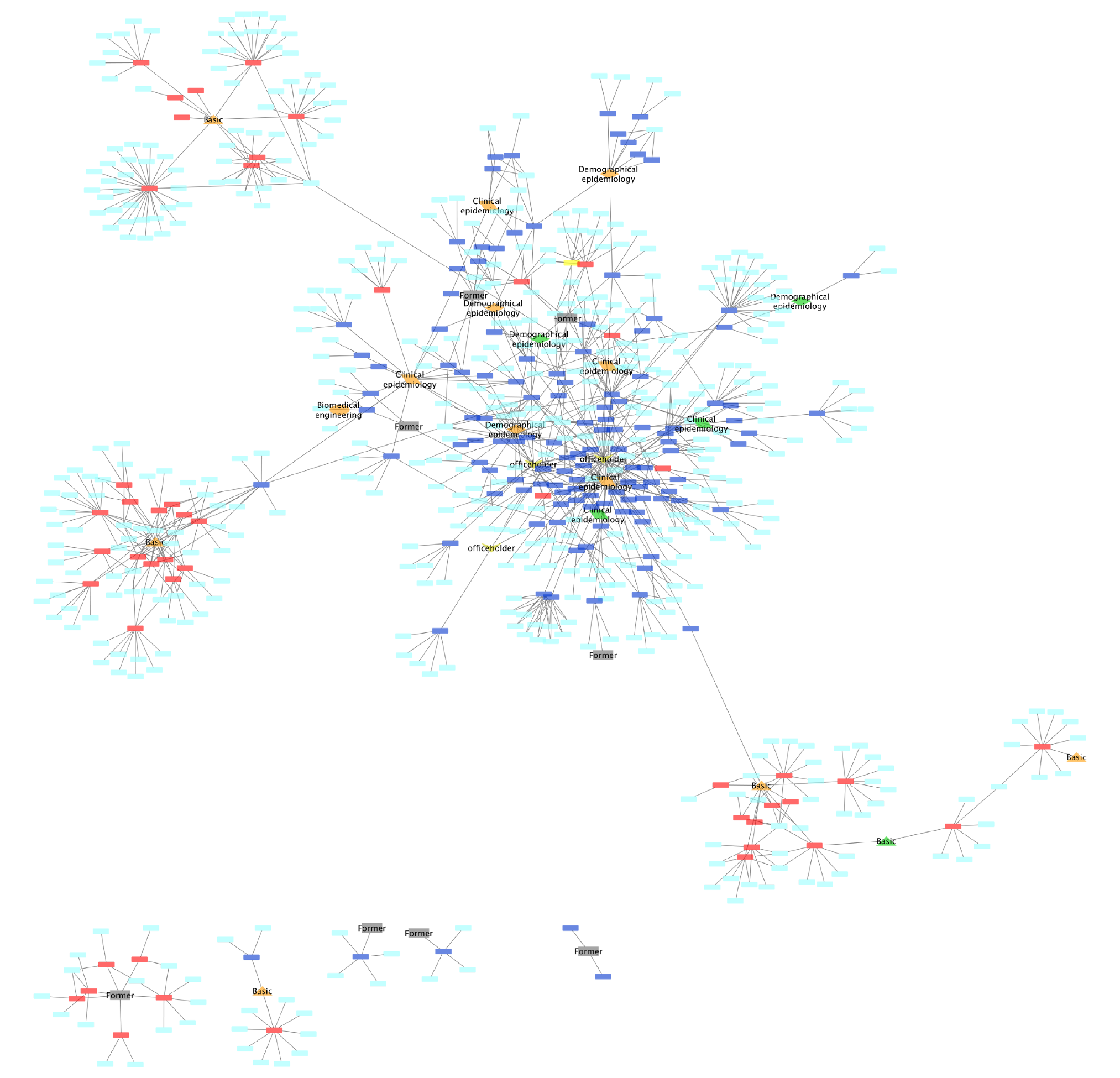}
  \caption{Bimodal network model of authors and papers. INGER researchers that belong to the department of basic research (sub-individual level) are colored in orange while INGER researchers that belong to the demographic and clinical departments are colored in green. Officeholder are colored in yellow. Former INGER researchers  are colored in gray. Authors from other institutions are in light blue. Basic research papers are in red while clinical papers are in dark blue.}
\end{figure}

\begin{figure}[H]
  \centering
\includegraphics[widht=15cm,height=22 cm]{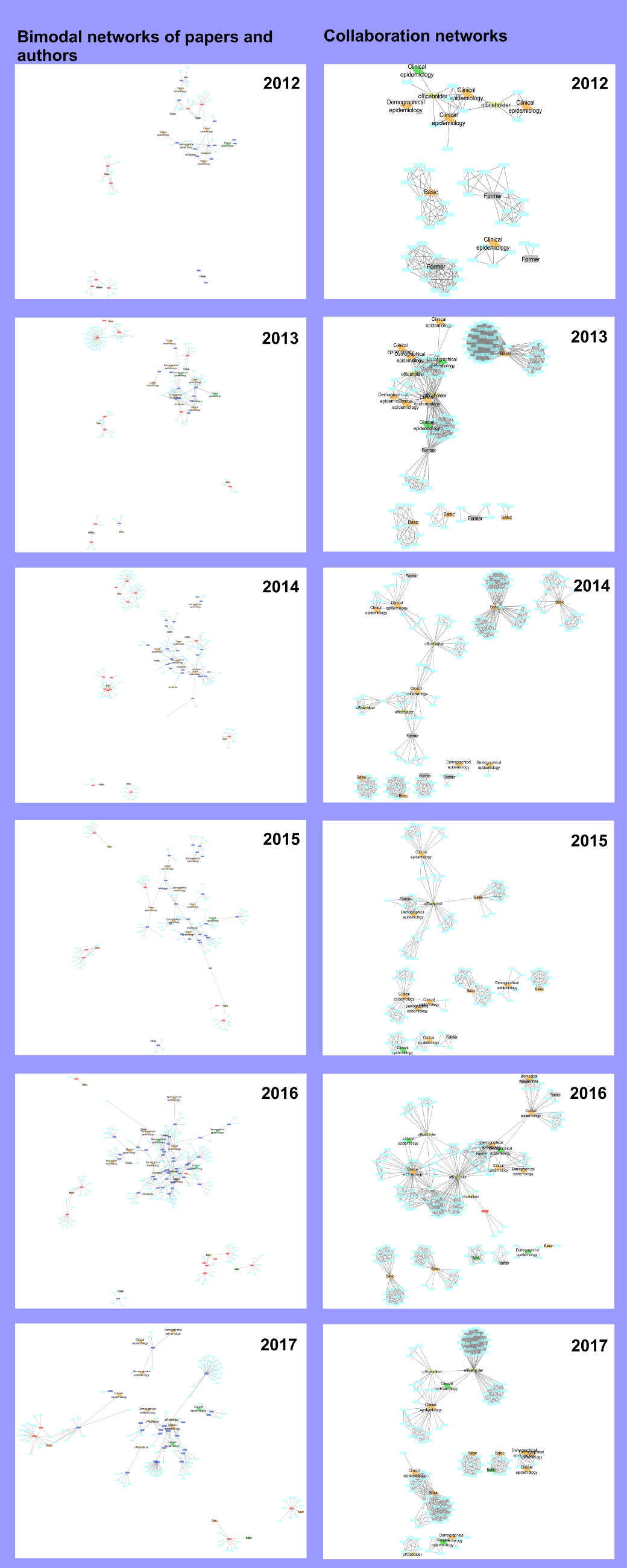}
  \caption{ Collaboration network model of INGER authors sliced by years (2012-2017). The authors were equally colored as in Figure 2. }
\end{figure}

\begin{figure}[H]
  \centering
  \begin{subfigure}[b]{.3\textwidth}
\centering  
\includegraphics[scale=0.28]{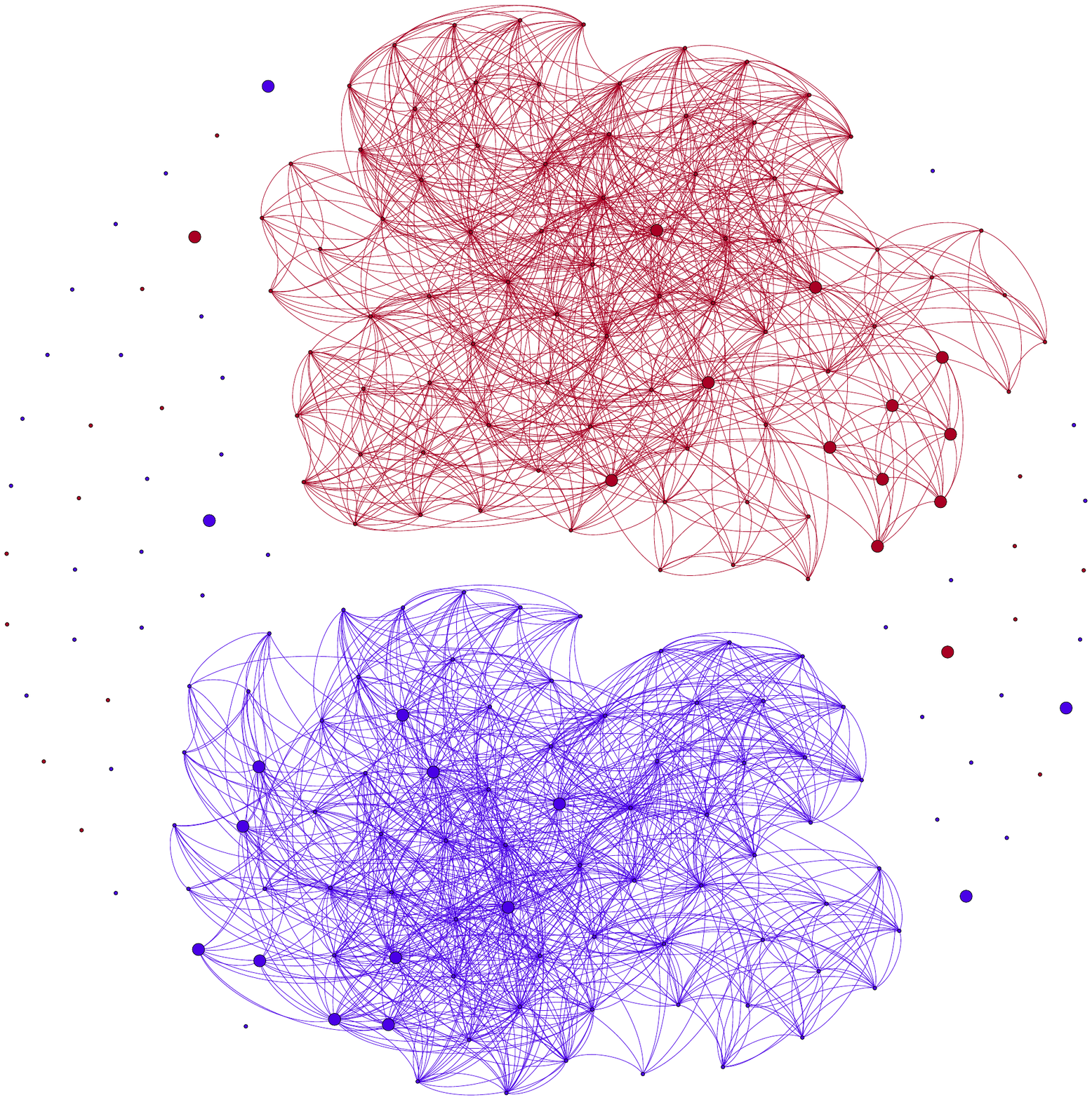}
\caption{INGER}
\end{subfigure}
  \begin{subfigure}[b]{.3\textwidth}
\centering
\includegraphics[scale=0.28]{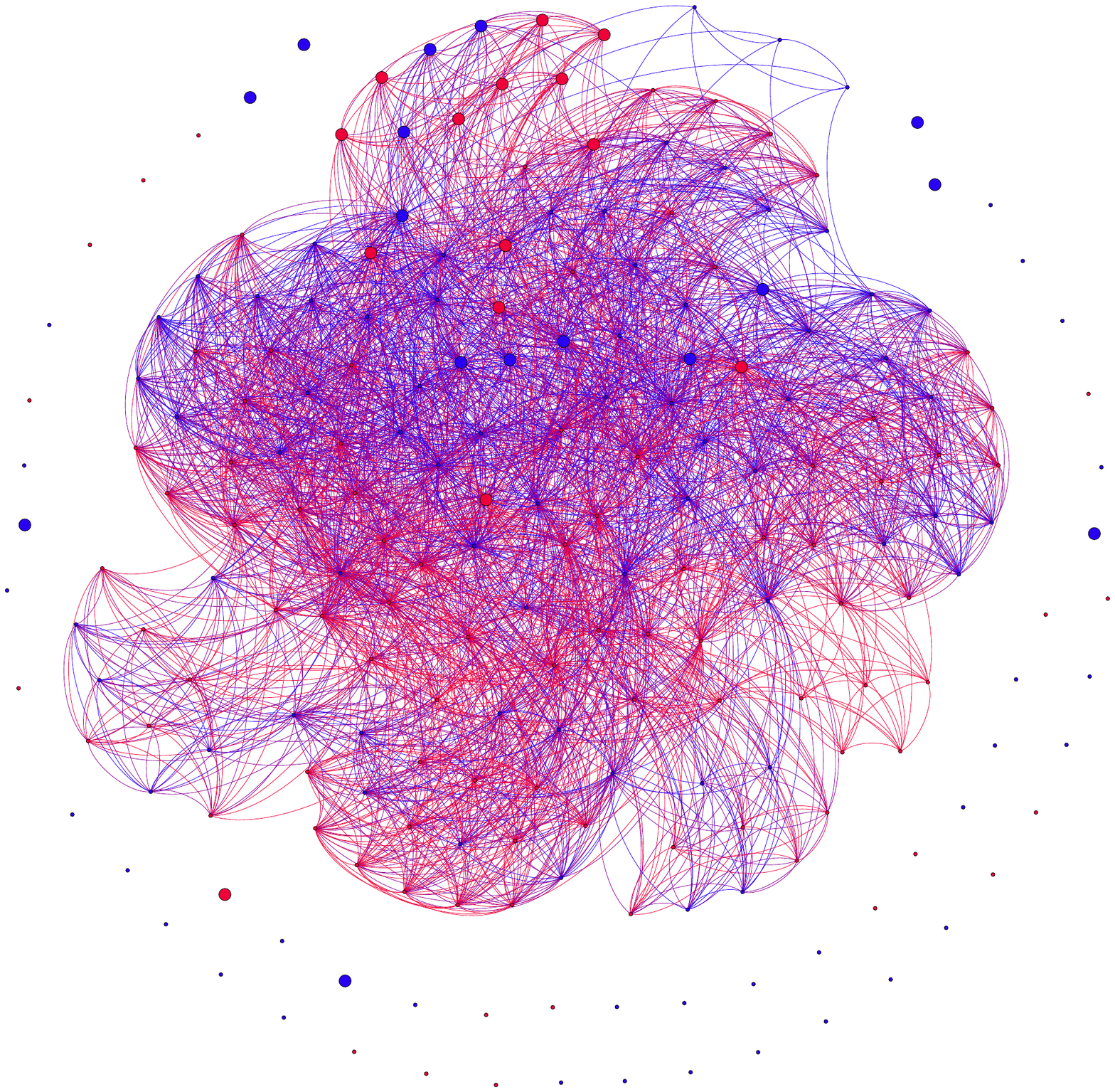}    
\caption{Full-Interdisciplinary}
  \end{subfigure}
\begin{subfigure}[b]{.3\textwidth}
\centering

\includegraphics[scale=0.28]{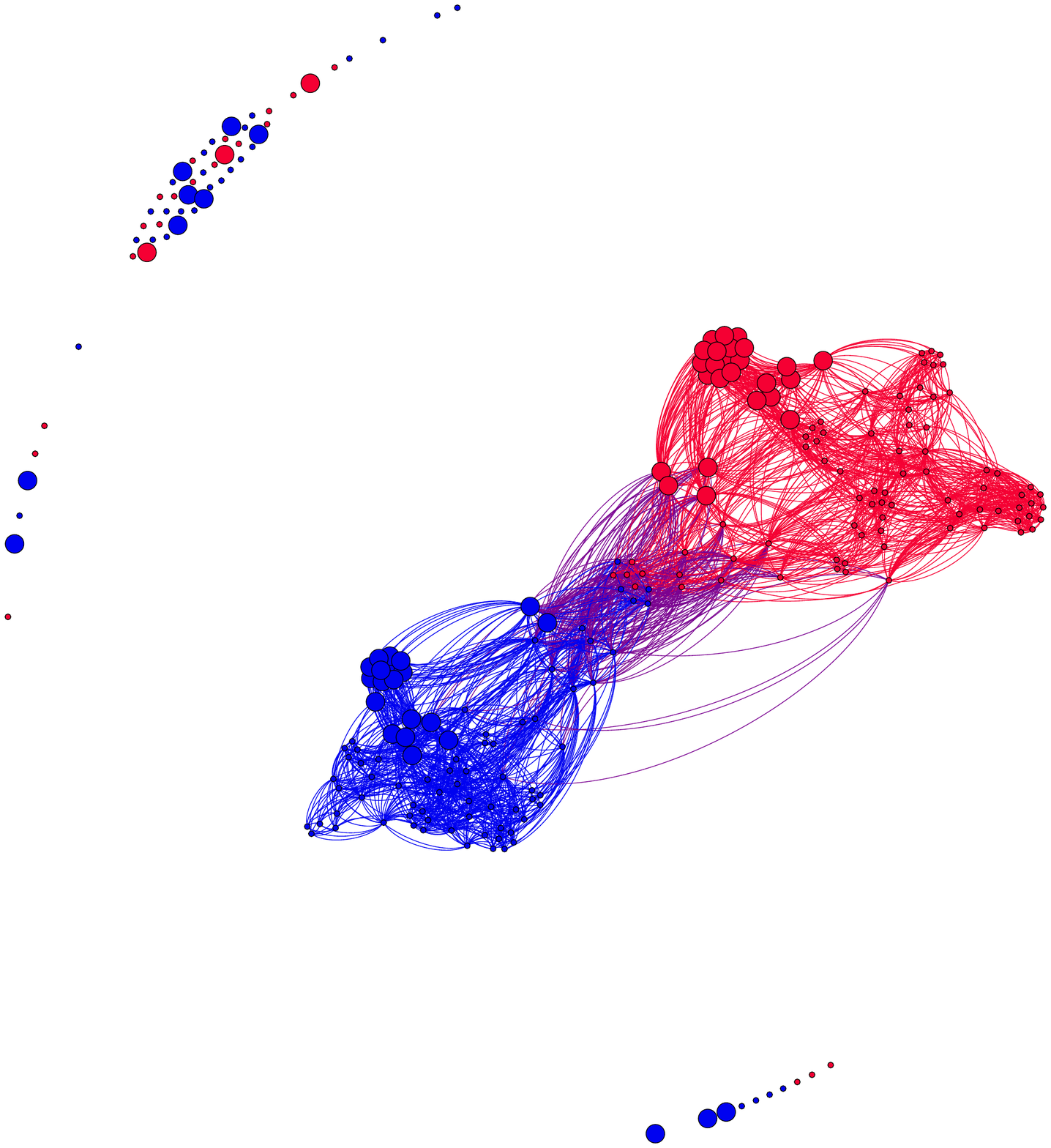}
\caption{Proposal}
  \end{subfigure}
\caption{Samples of collaboration networks corresponding to the three virtual experiments performed at three different values of homophily: a) 14\%, b) 46\% and c) 79\%. The experiments were performed in NetLogo through the inter-multidisciplinary team assembly model available here:  \url{http://modelingcommons.org/browse/one\_model/5676}}
\end{figure}

\begin{figure}[H]
  \centering
  \begin{subfigure}[b]{.8\textwidth}
\includegraphics[scale=0.27]{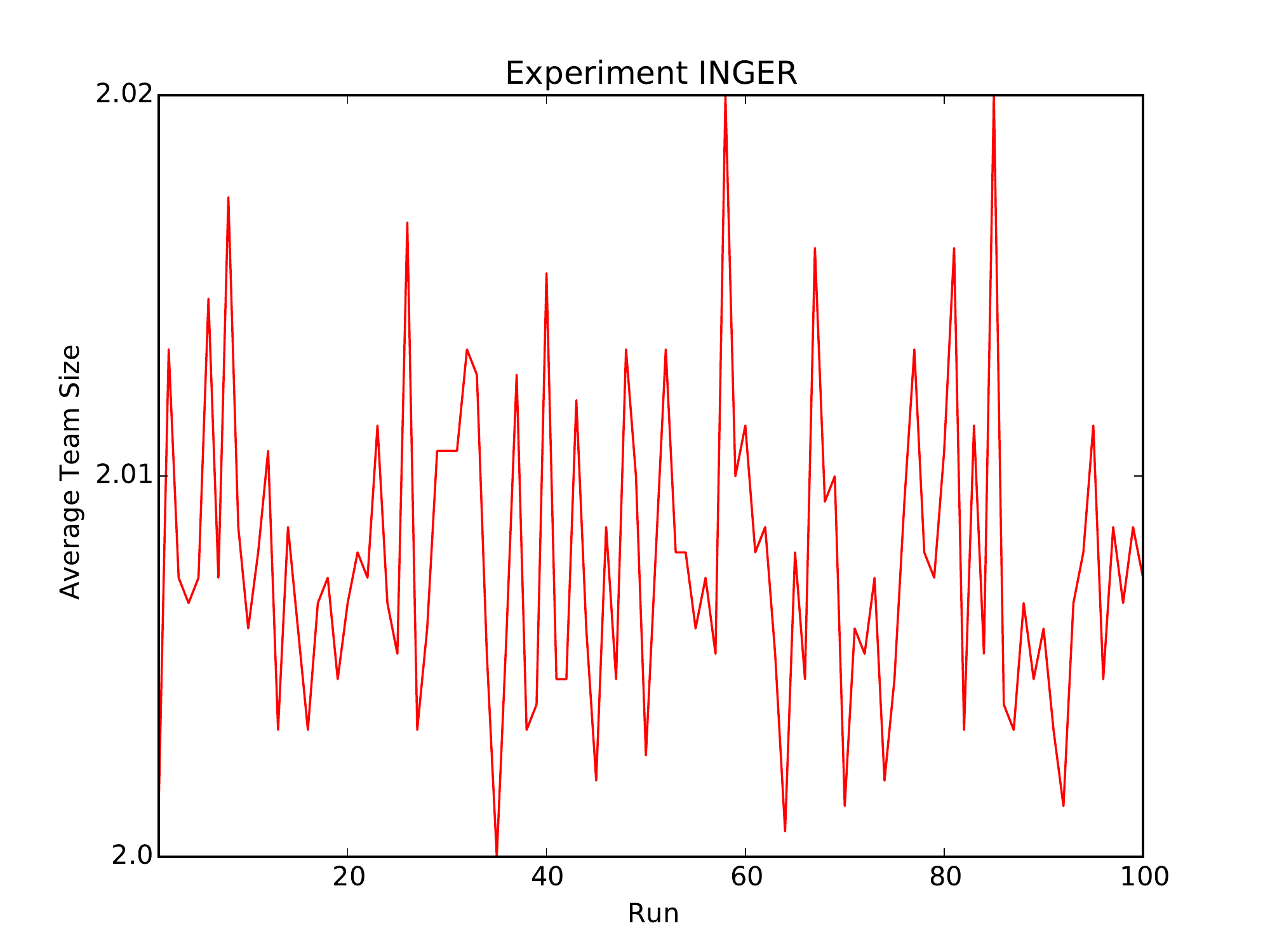}\includegraphics[scale=0.27]{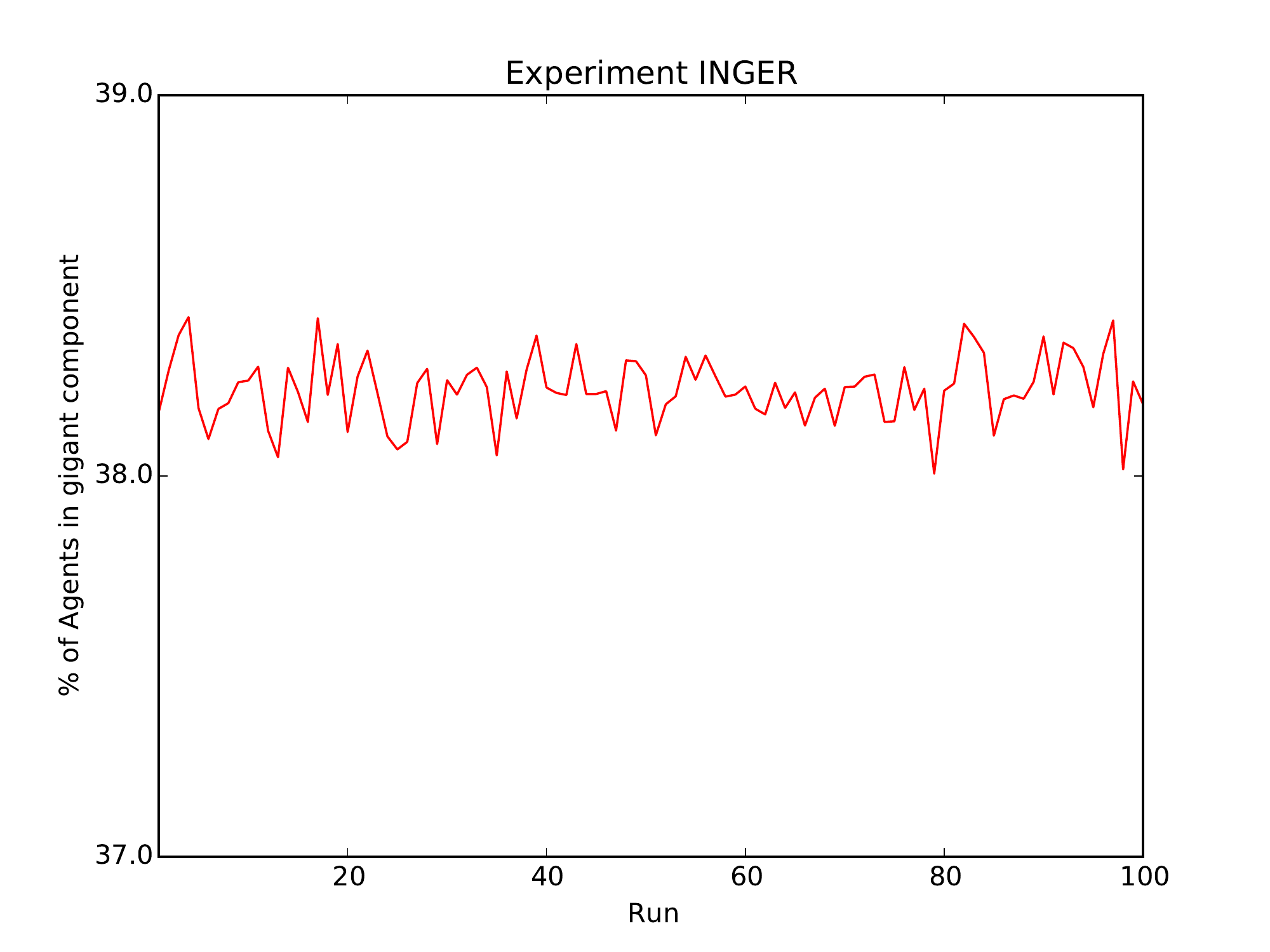}\includegraphics[scale=0.27]{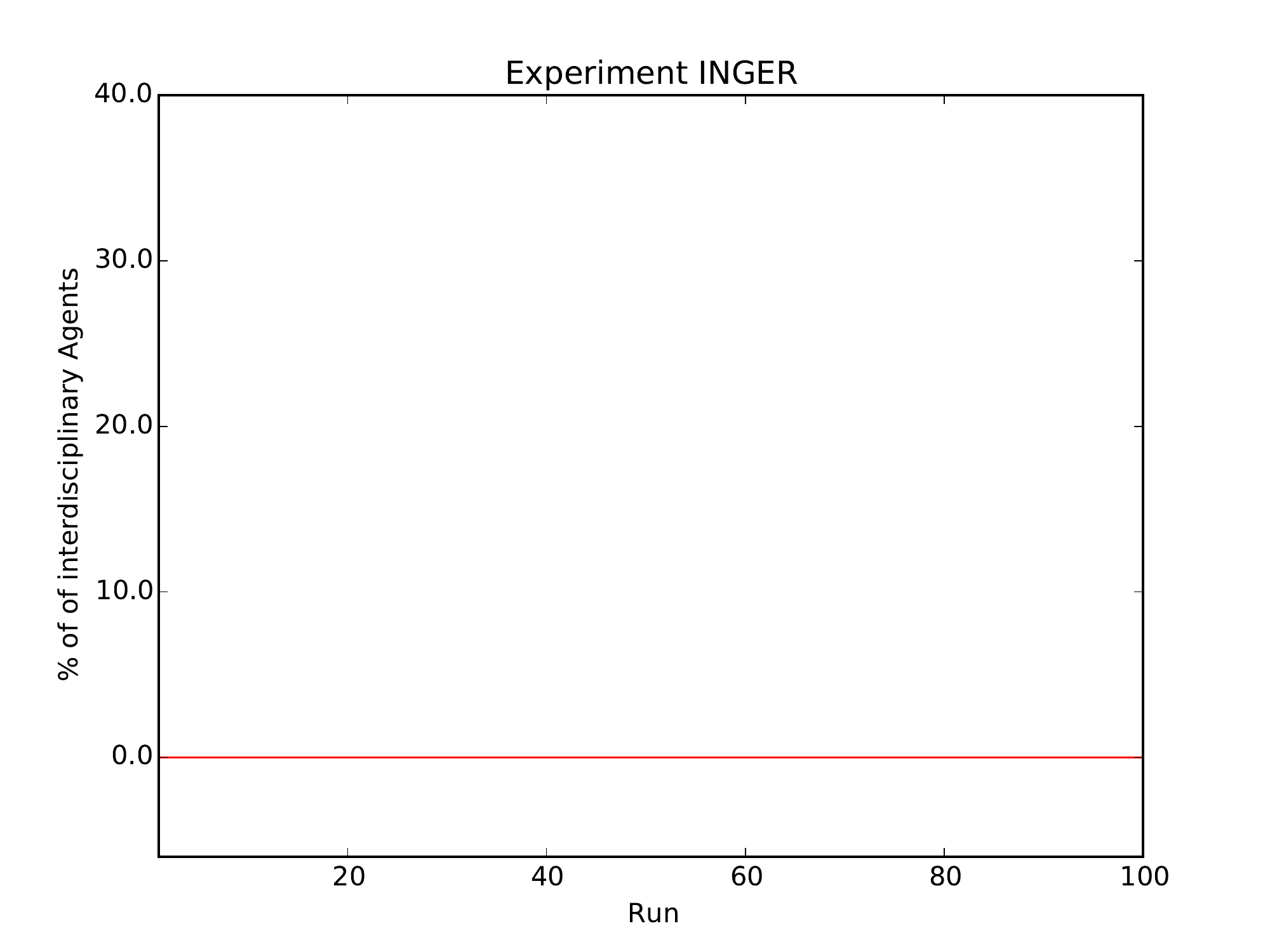}
\caption{INGER}
\end{subfigure}
  \begin{subfigure}[b]{.8\textwidth}
\includegraphics[scale=0.27]{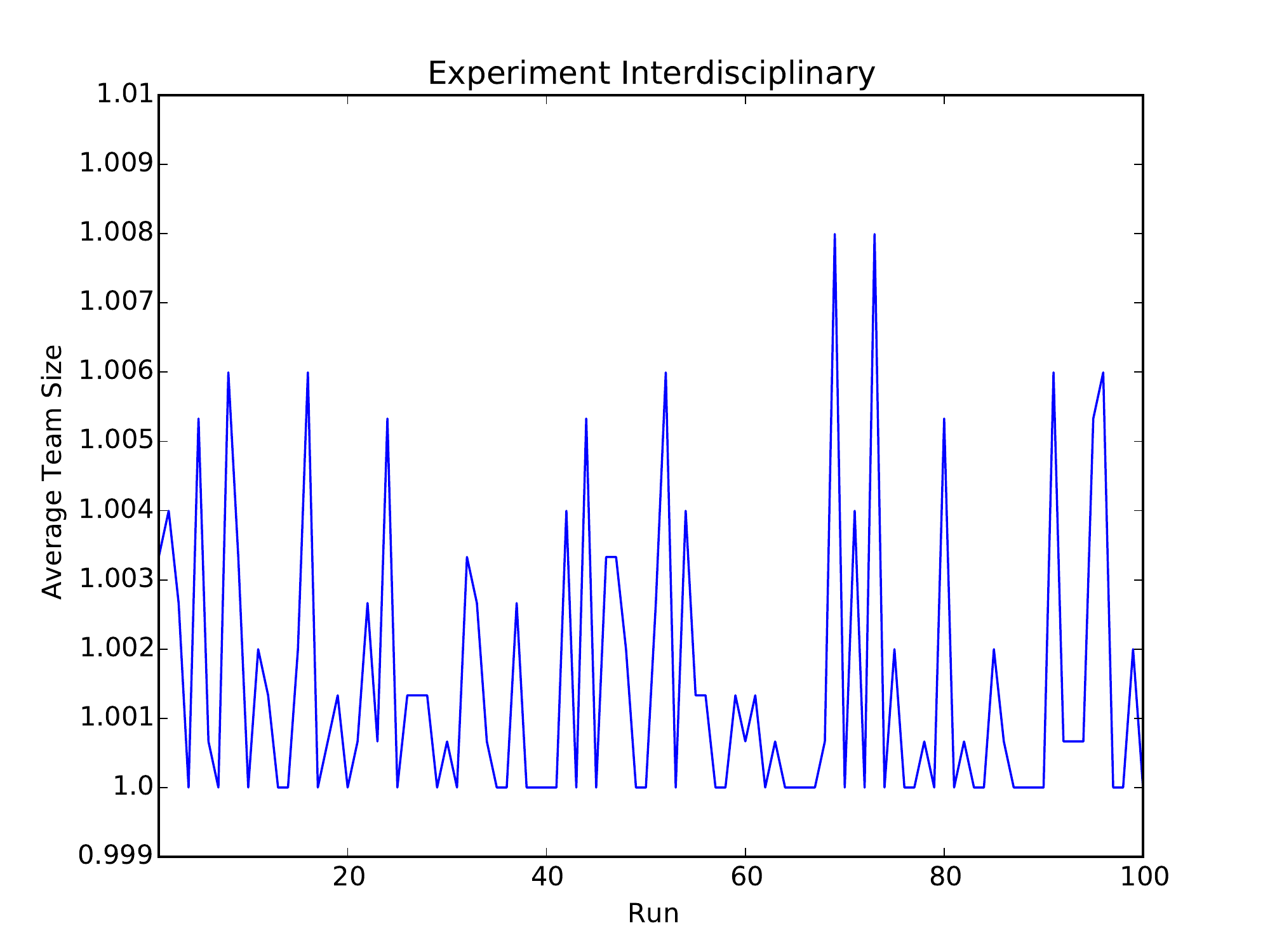}\includegraphics[scale=0.27]{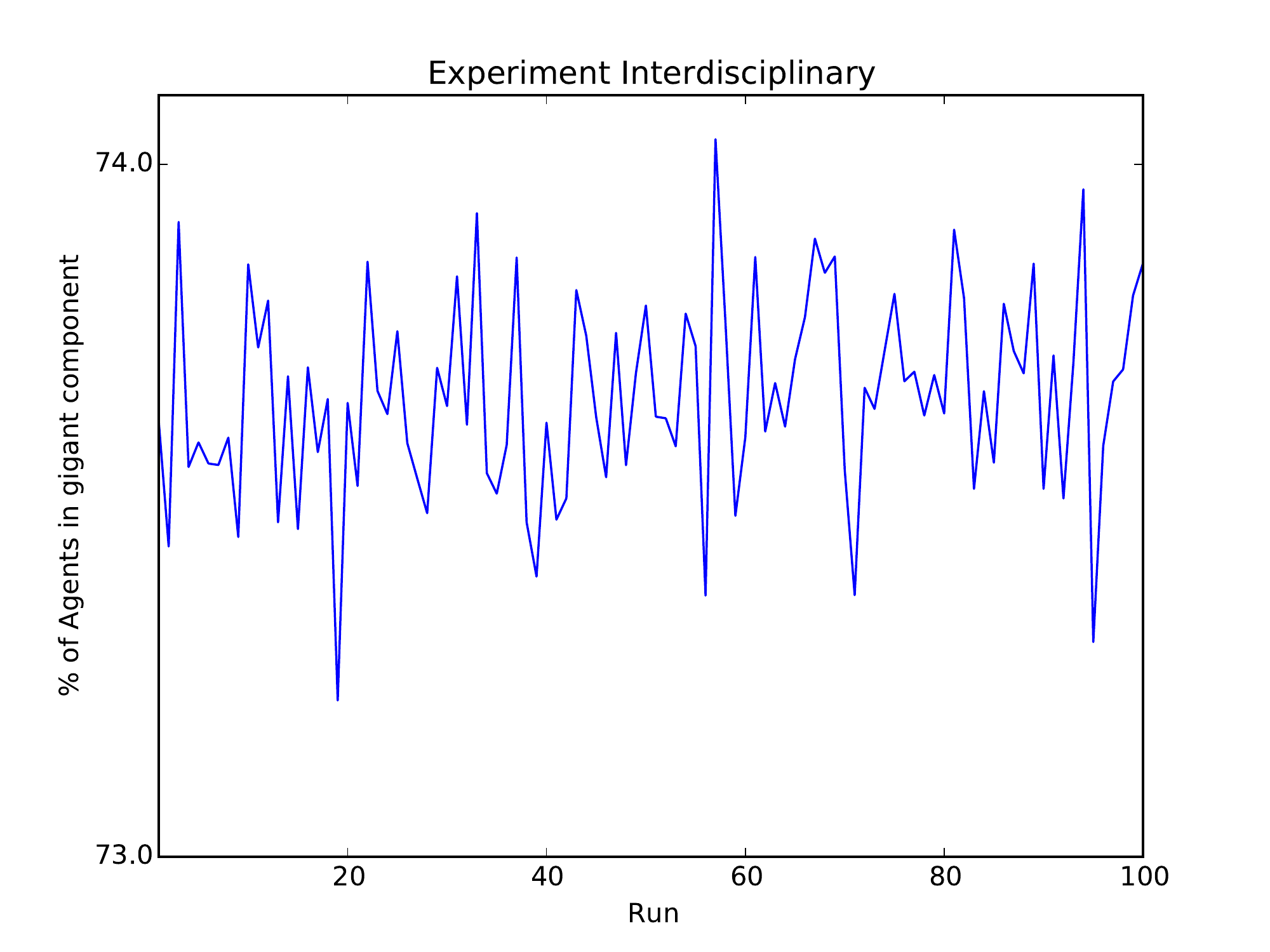}\includegraphics[scale=0.27]{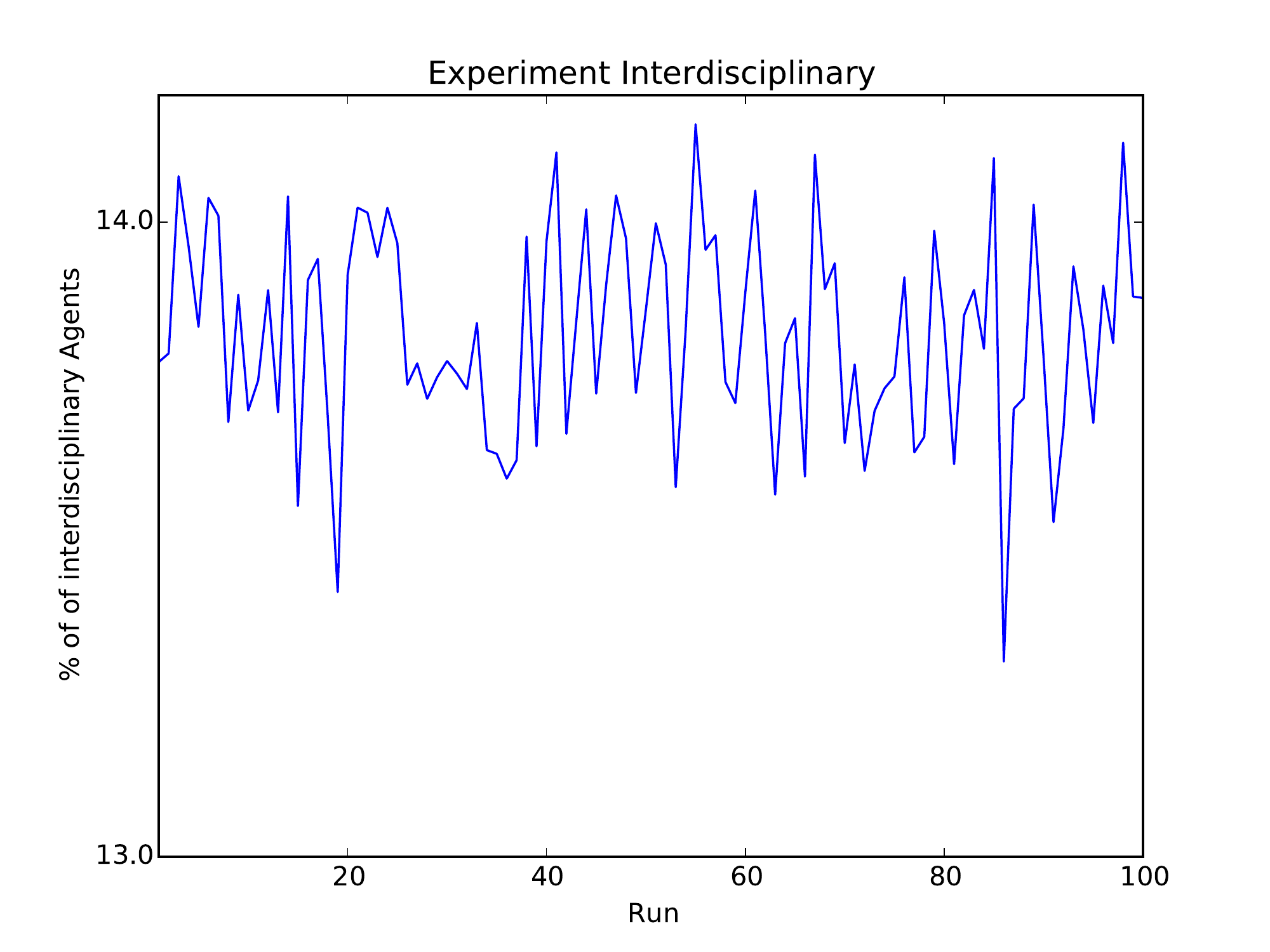}    
\caption{Full-Interdisciplinary}
  \end{subfigure}
\begin{subfigure}[b]{.8\textwidth}
\includegraphics[scale=0.27]{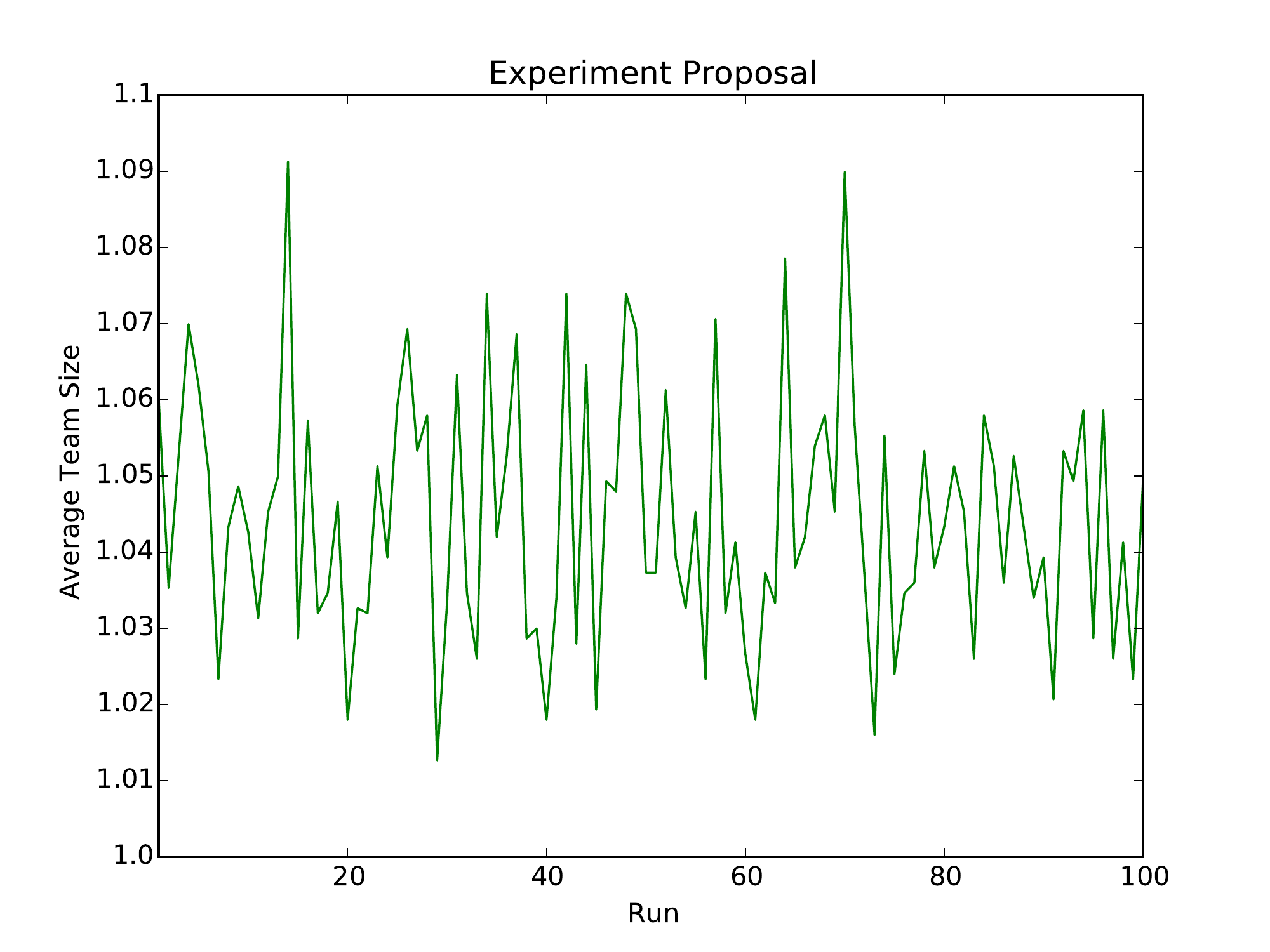}\includegraphics[scale=0.27]{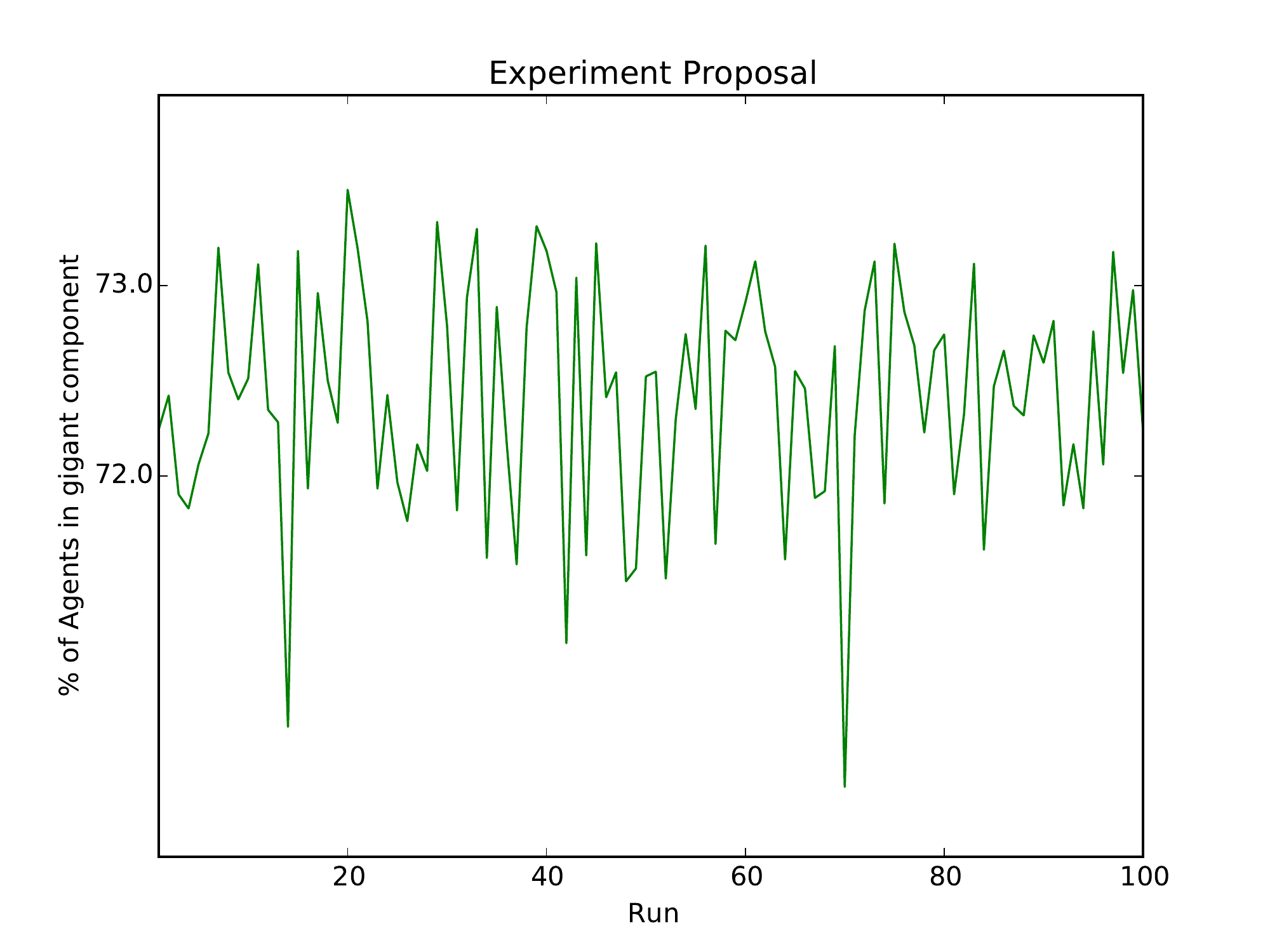}\includegraphics[scale=0.27]{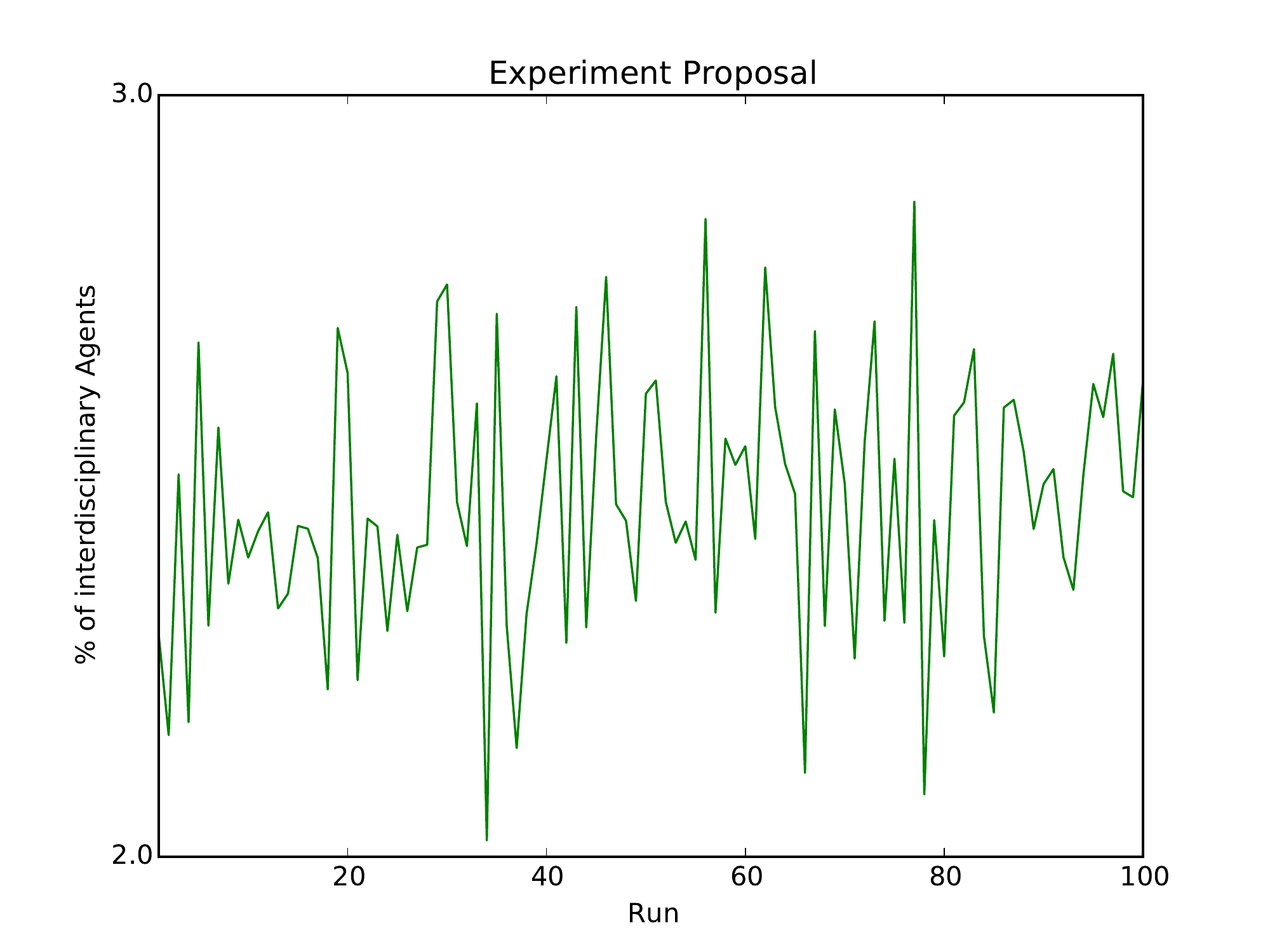}
\caption{Proposal}
  \end{subfigure}
\caption{  Average team size, percentage of agents in the the giant component and percentage of interdisciplinary agents (clinical researchers collaborating with basic researchers and vice versa) along the steps (just 100 ticks are showed). The plots corresponds with each of the three virtual experiments. }
\end{figure}

\end{document}